# Application and Extension of Mean-Field Theory such as SIR to Discuss the Non-Mean Field Problem of COVID-19

Hiroshi Isshiki, Institute of Mathematical Analysis (Osaka)

Masao Namiki, Former Vice President of Toshiba

Summary:

The concept of the effective infection opportunity population (EIOP) was incorporated into the SIQR model, and it was assumed that this EIOP would change with the spread of infection, and this was named as the effective SIQR model. When calculated with this model, the uninfected population S decreases with the passage of time. However, when the EIOP *N* increases because of any reason, the infection threshold becomes larger than 1. Even after the first wave seems to have subsided, the infection begins to spread again. Firstly, we find the curve of EIOP change so that the calculation result by this model matches the data of the first and second waves. Then, we use this curve to fit with only the data of the second wave alone, and the third wave is predicted. In the case of new coronavirus infection, there are various restrictions on data collection to identify individual coefficients of mathematical models, and the true value is almost unknown. Therefore, the discussion in this paper is only about data fitting for predictive calculation. Therefore, the simulation on the true value is not aimed. However, since the data of infected persons reflect the true values, the results of data fitting can be used for the prediction of infected persons, isolated care recipients, inpatients, and severely ill persons. They are useful for a qualitative understanding of infection. The idea of EIOP is important in the sense that it connects the mean-field and the non-mean field, but the existence of data is essential, and the theory alone cannot simulate the non-mean field. We have developed two methods for treating the non-mean field cases where we don't have enough data. We have briefly introduced them.

1. Introduction

Infection with the new coronavirus in Japan is less common than in Western countries, but it has also spread in Japan. Newly infected persons of the first wave which peaked in mid-April seemed to have converged once, but the second wave of infection spread from the beginning of June. It peaked out in early August, and by the end of August, it seemed that it was heading toward convergence. In Tokyo, on September 10th, the alert level of the infection status due to the four stages of the new coronavirus was finally lowered from the most serious "infection is spreading" to the second level " Be wary of re-spreading the infection. "

During this period, public institutions did not accurately predict inpatients and severely ill people, and in some prefectures, there was a serious concern that there would be a shortage of beds. The New-type Coronavirus Infection Mathematical Model Study Group (https: note.com/antivirus) uses SIQR models [1] to calculate the predictions of newly infected persons, hospitalized persons, and severely ill persons for major prefectures and publicized them on a homepage. It was released 7 times from August 4th as "Corona Forecast (COVID-19 Forecast) [2]". This "corona forecast" was temporarily suspended on August 30 due to the peak out of the second wave, but it is evaluated that the prediction accuracy was practically sufficient. I think most of the current concerns are whether the third wave will come or not, and if so, when the infection will begin, but no one seems to have a valid predictive calculation. Some people say, "It will come around autumn," or "It's serious if it comes around winter and overlaps the flu," but this is a prophecy, not a scientific prediction.

In order to make a prediction, it is necessary to accurately simulate (regress) the current infection status on the data. In other words, once the past trends are learned by trend analysis, it is a common method of prediction to extend the trends to the future. So what should we use for the curve that regresses and extends the data to the future? In addition to mathematical rationality, this selection also requires understanding such as the nature of the event to be predicted. The case of COVID-19 infection is complicated because it has both biological and sociological aspects. In the "Corona Forecast", a logistic curve, which is a growth curve, and Otagaki's SIQR mathematical model [1] are used together with this curve. However, since both are first-order differential equations, the diffusion speed can be simulated, but the oscillating phenomenon cannot be simulated. In order to continuously regress the first wave to the second wave, it is necessary to improve the equation. Therefore, by incorporating the concept of effective infection opportunity population (EIOP) [3] into Otagaki's SIQR mathematical model (hereinafter referred to as SIQR model), we devised a mathematical model called the effective SIQR model. This made it possible to regress two or more waves in one calculation. This is introduced below.

With the introduction of EIOP, the phenomenon of the non-mean field can be treated by the equation of mean-field to some extent, but the existence of data is indispensable for that purpose, and it cannot be used for theoretical simulation. As can be read from the results of a simulation using on the cellular automaton [6], the SIR-type equation for the mean-filed phenomena should be extended to be able to discuss the non-mean-field phenomena. Two methods have been devised to avoid such problems [7,8]. One is a method of theoretically incorporating the social effect on the infectious phenomenon, and the other is a spatially discrete SIR type equation for dealing with the non-mean-field so that the phenomenon of the non-mean field can be dealt with directly. We also discussed both of these and obtained useful results.

## 2. Application of mean-field Equation to Non-mean-field
### 2.1 Concept of Effective Infection Opportunity Population (EIOP)

The SIQR model (given by Eq. (1) not including $dN(t)/dt$) is extended. In it, the concept of the Effective Infection Opportunity Population (EIOP) is important. In the SIR model, there is a population conservation law, which includes uninfected or susceptible persons $S$ (Susceptible), infected persons $I$ (Infected), and recoverers (+ dead) $R$ (Recovered). Then, the total population $N$ satisfies the population conservation law $N = S + I + R$. In the SIQR model, the infected persons have divided into the in-the-city infected-persons $I$ and the isolated infected-persons $Q$ (Quarantined). Then, $N$ satisfies $N = S + I + Q + $ .

The idea of EIOP was initially the idea that was reached by repeating data fitting by changing the coefficients of the SIQR model in various ways. When revisiting the daily trend of newly infected people throughout Japan, we found that it was unreasonable to use Japan's total population of 127 million for the total population $N$. The reason is as follows. The infection in the first wave spread and converged during the period of about two months.

The population that has the opportunity to be frequently exposed to infected people during the period is extremely small compared to the total population of Japan even if the people were without self-restraint. Currently, the SIR model that can be used to analyze new coronavirus infections is the mean-field model. The phenomenon that actually occurs is a non-mean field that spreads from the cluster. Populations far from the cluster are not involved in infection and must be excluded from the calculation (time exclusion). In addition, even in the vicinity of the cluster, self-restraint behaviors such as wearing masks are not involved in infection and must be excluded from the calculation (self-restraint exclusion). Instead of using $N$ as a premise value, it should be considered as an unknown quantity whose value is determined by data fitting. It is not rational to use Tokyo's population of 13 million for $N$ for the same reason. Then, how should the population $N$ used for the mean-field mathematical model such as the SIR system be determined?

Let's think from another point of view. In the mean-field theory, infected and susceptible people are in contact everywhere, creating the contradiction that infection occurs even among populations that the infection has not reached from the cluster.

When calculating the non-mean field due to the spread of one infected person into the population space by the calculation of cellular automaton [6] or particle method, if the basic reproduction number is 1 or more, the infection spreads endlessly. However, the spreading speed is finite. By the time the infection spreads to the whole, as shown in Fig. 1, uninfected areas will be created (time effect), and the excluded population will occur.

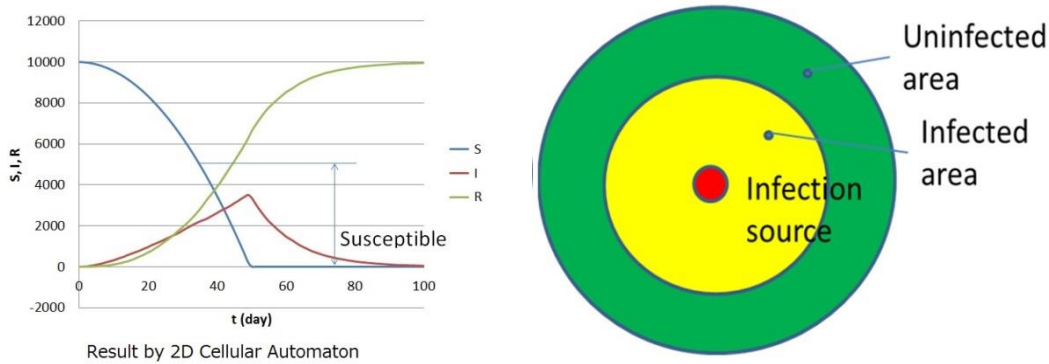

Fig. 1 Time effect and excluded population.

Fig. 2 is a conceptual diagram of the EIOP $N$ used in the calculation. The part surrounded by the broken line in Fig. 2 is the population exposed to the infected person. Regardless of the size of the group, if we call it a cluster, we should think that the population $N$ represented by the mathematical model of the SIR system is the total number of people in this cluster. That is, population $N$ is the sum of the population of only those involved in the infection. Therefore, those who are steadily implementing masks and three-closeness avoidance as described above are outside this population because they are not involved in infection. As the spread of infection continues, the spreaders of viruses, mainly unaffected infected people, increase, so the population $N$ involved in infection increases as shown from the left figure to the right figure in Fig. 2.

As mentioned above, it is almost impossible to physically count this number of people, so in the calculation, $N$ is also an explanatory variable like other coefficients of the mathematical model. Calculate by curve-fitting so that the calculation result matches the data of newly infected persons in one day. It is similar to the so-called AI machine learning method. The $N$ required in this case is the average over the period of infection.

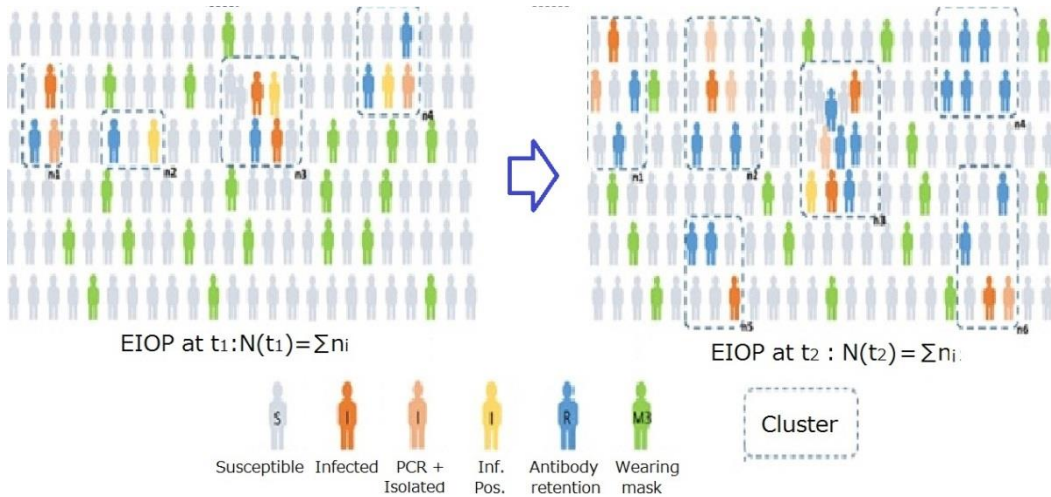

Fig. 2 Concept of Effective Infection Opportunity Population (EIOP).

Table 1 SIQR model and calculated values by logistic curve regression.

| Area | Pop./10³ | Wave | SIQR parameters ($q$=0.2, 1/$\gamma$=9 days) | | | | | | Parameters in logistic-curve-fitting | | | | H'=b/N | Ro' | Data-term |
| | Pop. dens. | | $N$ | $N/N_{tot}$ | $\beta$ | $\beta'=\beta/N$ | $q'$ | $R^2$ | $a$ | $b$ | $b/N_{tot}$ | $R^2$ | | | |
|---|---|---|---|---|---|---|---|---|---|---|---|---|---|---|---|
| Japan | 127,095 | 1st | 41,263 | 0.03% | 0.470 | 1.14E-05 | 0.0135 | 0.88 | 0.135 | 15,841 | 0.01% | 0.87 | 38% | 1.62 | 3/10-5/21 |
| | 341 | 2nd | 232,062 | 0.18% | 0.400 | 1.72E-06 | 0.0002 | 0.88 | 0.085 | 59,633 | 0.05% | 0.87 | 26% | 1.35 | 5/11-9/11 |
| Hokkaido | 5,381 | 1st | 1,985 | 0.04% | 0.621 | 3.13E-04 | 0.2280 | 0.67 | 0.124 | 962 | 0.02% | 0.61 | 48% | 1.94 | 3/10-5/21 |
| | 69 | 2nd | 3,311 | 0.06% | 0.376 | 1.14E-04 | 0.0000 | 0.34 | 0.087 | 618 | 0.01% | 0.29 | 19% | 1.23 | 7/6-9/12 |
| Saitama | 7,267 | 1st | 1,945 | 0.03% | 0.728 | 3.74E-04 | 0.3300 | 0.73 | 0.145 | 971 | 0.01% | 0.72 | 50% | 2.00 | 3/10-5/21 |
| | 1,913 | 2nd | 14,245 | 0.20% | 0.611 | 4.29E-05 | 0.3830 | 0.76 | 0.073 | 3,430 | 0.05% | 0.75 | 24% | 1.32 | 5/11-9/11 |
| Chiba | 6,223 | 1st | 2,201 | 0.04% | 0.475 | 2.16E-04 | 0.0000 | 0.49 | 0.147 | 865 | 0.01% | 0.49 | 39% | 1.65 | 3/10-5/21 |
| | 1,206 | 2nd | 11,466 | 0.18% | 0.420 | 3.66E-05 | 0.0810 | 0.65 | 0.080 | 2,596 | 0.04% | 0.67 | 23% | 1.29 | 5/11-9/11 |
| Tokyo | 13,515 | 1st | 15,843 | 0.12% | 0.436 | 2.75E-05 | 0.0000 | 0.65 | 0.127 | 5,108 | 0.04% | 0.61 | 32% | 1.48 | 3/10-5/21 |
| | 6,168 | 2nd | 85,907 | 0.64% | 0.411 | 4.78E-06 | 0.0790 | 0.77 | 0.072 | 18,272 | 0.14% | 0.76 | 21% | 1.27 | 5/11-9/11 |
| Kanagawa | 9,126 | 1st | 3,346 | 0.04% | 0.465 | 1.39E-04 | 0.0289 | 0.46 | 0.130 | 1,209 | 0.01% | 0.44 | 36% | 1.57 | 3/10-5/21 |
| | 3,778 | 2nd | 28,056 | 0.31% | 0.374 | 1.33E-05 | 0.0000 | 0.70 | 0.066 | 5,291 | 0.06% | 0.69 | 19% | 1.23 | 5/11-9/11 |
| Osaka | 8,839 | 1st | 4,163 | 0.05% | 0.484 | 1.16E-04 | 0.0023 | 0.76 | 0.147 | 1,704 | 0.02% | 0.75 | 41% | 1.69 | 3/10-5/21 |
| | 4,640 | 2nd | 28,005 | 0.32% | 0.412 | 1.47E-05 | 0.0006 | 0.74 | 0.107 | 7,532 | 0.09% | 0.72 | 27% | 1.37 | 5/11-9/11 |
| Hyogo | 5,535 | 1st | 1,432 | 0.03% | 0.502 | 3.51E-04 | 0.0023 | 0.46 | 0.117 | 705 | 0.01% | 0.42 | 49% | 1.97 | 3/10-5/21 |
| | 659 | 2nd | 6,089 | 0.11% | 0.421 | 6.91E-05 | 0.0007 | 0.83 | 0.114 | 1,749 | 0.03% | 0.85 | 29% | 1.40 | 5/11-9/11 |

In this way, Table 1 shows the coefficients (explanatory variables) of the SIQR model and the logistic curve obtained by curve-fitting the data of newly infected persons in the first and second waves of each of the major prefectures. $N$ in the "SIQR model calculation" group in Table 1 is the EIOP. The infection rate $\beta'$ and the test isolation rate $q'$ are also calculated at the same time as explanatory variables. $R^2$ in the same column is the coefficient of determination that indicates the goodness of fitting when a curve fit is performed. The $a$ in the "logistic curve regression" group indicates the steepness of the curve. $b$ is the arrival point of the S-shaped curve indicated by the logistic curve, that is, the cumulative value of infected persons. The second wave is still in the process of converging the infection, but the scale of the first wave and the second wave in each region can be compared by the value of $b$. $R^2$ of the same group is the coefficient of determination of logistic curve regression. $H = b / N$ shown in the column next to it is the ratio of the EIOP $N$ to the cumulative infection person $b$, and corresponds to the herd immunity threshold in the EIOP of each wave. If this herd immunity threshold is $H'$, the basic reproduction number $Ro'$ for this $H'$ can be calculated.

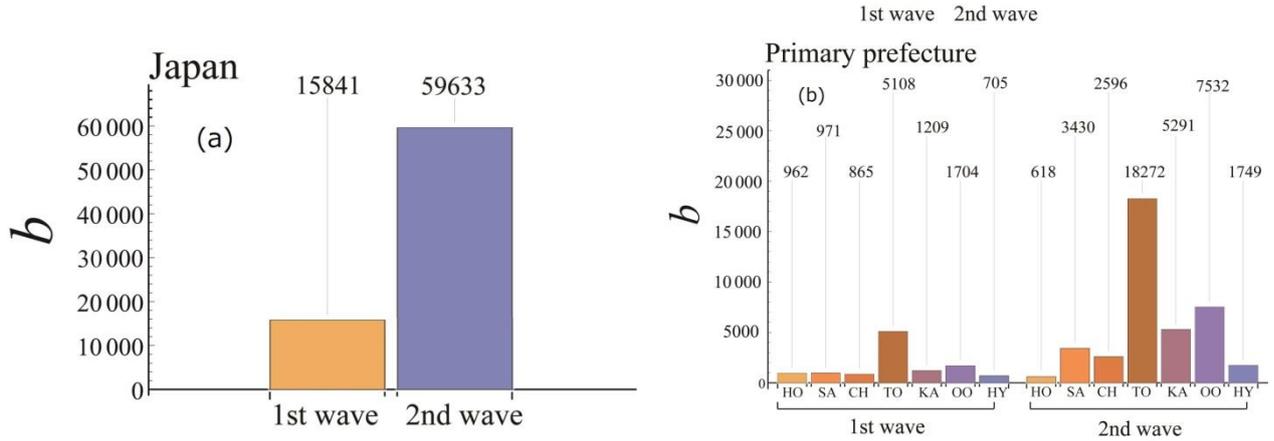

Fig. 3 Infection scale in major prefectures

(HO: Hokkaido, SA: Saitama, CH: Chiba, TO: Tokyo, KA: Kanagawa, OO: Osaka, HY: Hyogo).

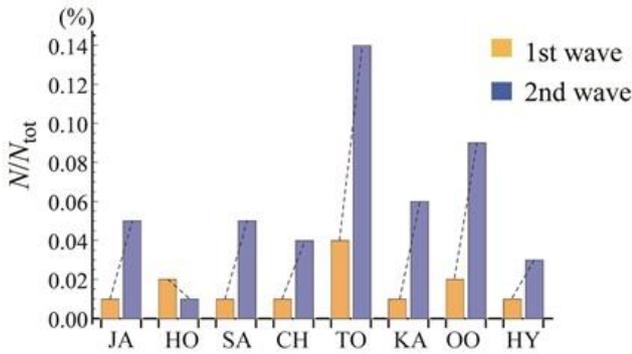
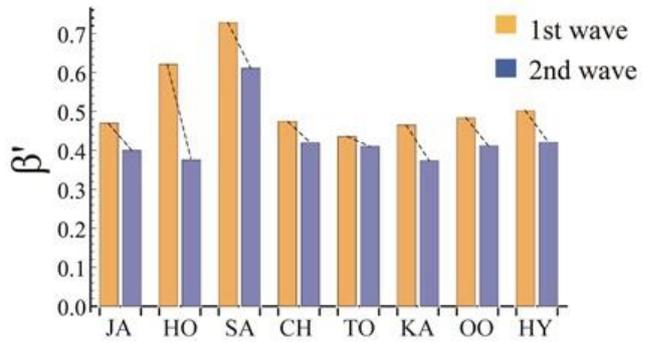

Fig. 4 Comparison of EIOP.    Fig. 5 Infection rate β' at 1st and 2nd waves.

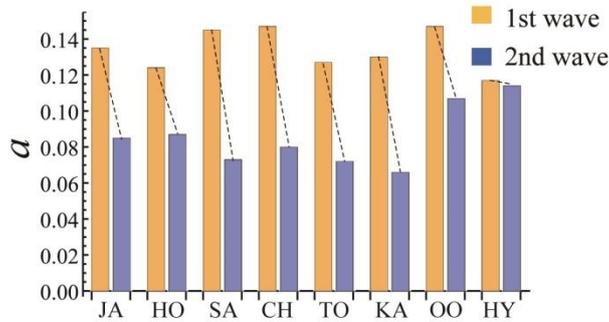

Fig. 6 Coefficient $a$ at 1st and 2nd waves.

Fig. 3 shows the comparison of the infection scale between the 1st and 2nd waves in the whole country and major prefectures. Fig. 4 is a graph of the EIOP $N$ in Table 1 divided by the actual population of the area on the vertical axis. Depending on the scale of the infection, it is larger in the second wave, but not necessarily the same. Fig. 5 is a graph with the infection rate $β'$ on the vertical axis, and $β'$ tends to decrease in the second wave. Although not described here, the difference between the first and second waves lies in the size of the infection. However, compared to the first wave, the proportion of severely ill and dead people in the second wave is smaller than expected from the scale of newly infected people. There is also the view that it can be explained by the age structure of the infected person and that the virus has been mutated to be attenuated. Since the highly virulent virus that causes aggravation and death is isolated together with the patient, it is presumed that the statistical probability will decrease. In addition, since there is a relatively high probability that a virus with weak infectivity will pass the infection period before infecting others, it seems reasonable to think that what remains in the city is a weakly virulent and highly infectious virus. Based on this hypothesis, the infection rate $\beta$ of the second wave and the basic reproduction number $Ro'$ should be larger than those of the first wave. However, $\beta'$ and $Ro'$ shown in Table 1 show the opposite tendency, suggesting that the influence of social factors is greater. Since $\beta'$ is also obtained in combination with $N$ and other coefficients, it is premature to conclude with this data, but it is anxious.

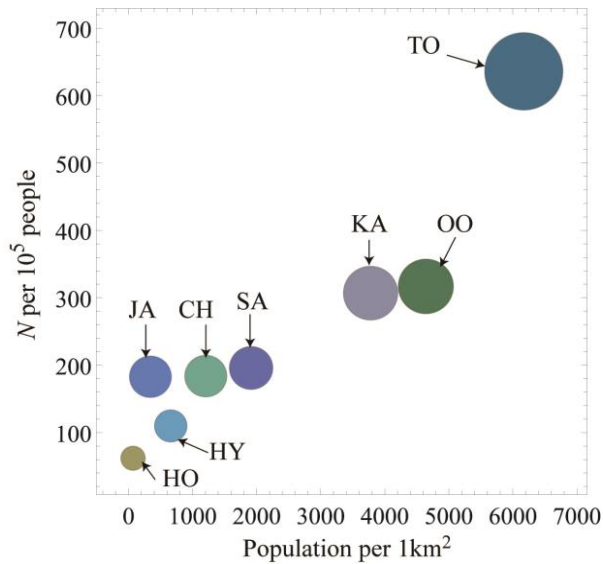

Fig. 7 EIOP $N$ in major prefectures (second wave)

(JA: Japan, HO: Hokkaido, SA: Saitama, CH: Chiba, TO: Tokyo, KA: Kanagawa, OO: Osaka, HY: Hyogo).

Fig. 6 compares the coefficient $a$ of the logistic curve regression between the first wave and the second wave, and it can be seen that the second wave tends to be smaller and the rise of the mountain is gentler. Fig. 7 shows the graph of the EIOP $N$ of the second wave in Table 1 converted to the value per 100,000 people on the vertical axis and the population density on the horizontal axis. The size of the bubble is the scale of infection. As a matter of course due to the nature of infection, the EIOP $N$ per 100,000 people increases in proportion to the population density. This can be seen as indicating that the smaller the population density, the larger the time exclusion.

**2.2 Effective Infection Opportunity Population (EIOP) that changes with the spread of infection**

Fig. 8 shows the data of newly infected persons in Tokyo and the regression of the first wave and the second wave by logistic curves, respectively. The blue and gray horizontal lines in the figure are the EIOP of the first and second waves obtained by curve-fitting with the SIQR model, respectively. With the conventional SIR model equation or growth curve (logistic curve or Gompertz curve), only one wave can be regressed, so prediction calculation cannot be performed until a while after the infection starts. 'For a while' means that useful predictive calculations can be made when the curve reaches the inflection point just before the peak of the infected mountain. In addition, as mentioned above, in the conventional calculation method, when one wave converges, the number of newly infected persons in the calculation decreases toward 0 and does not increase again. Even if the infection rate or healing period is changed, the shape of the mountain of infection only changes, and it is not possible to simulate infection after the next wave.

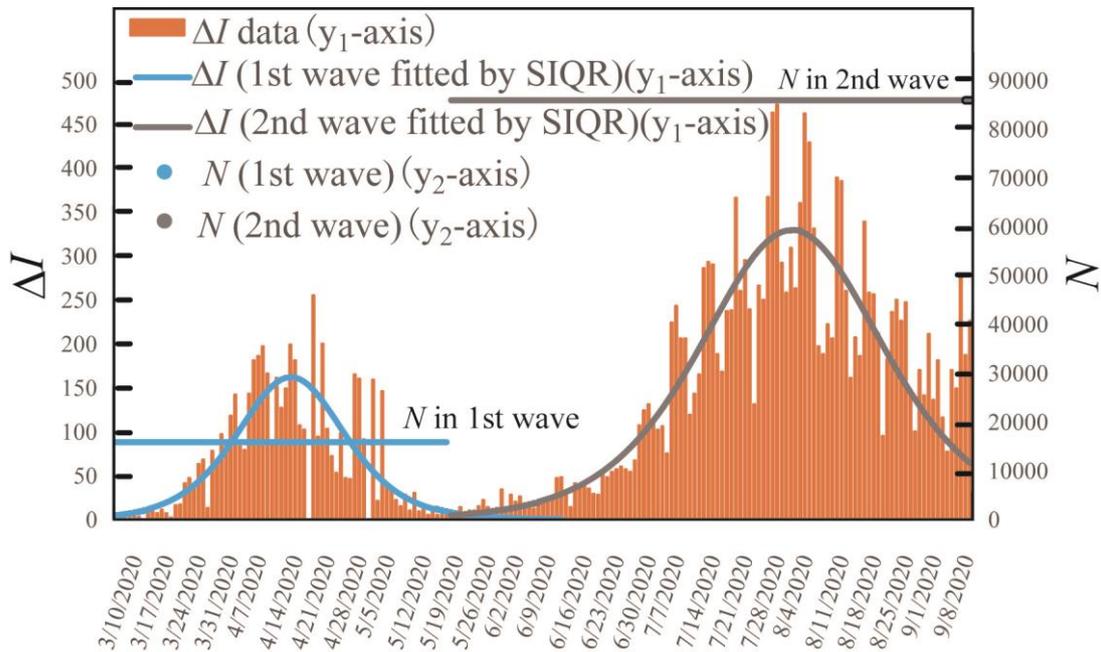

Fig. 8 Regression by conventional SIQR model.

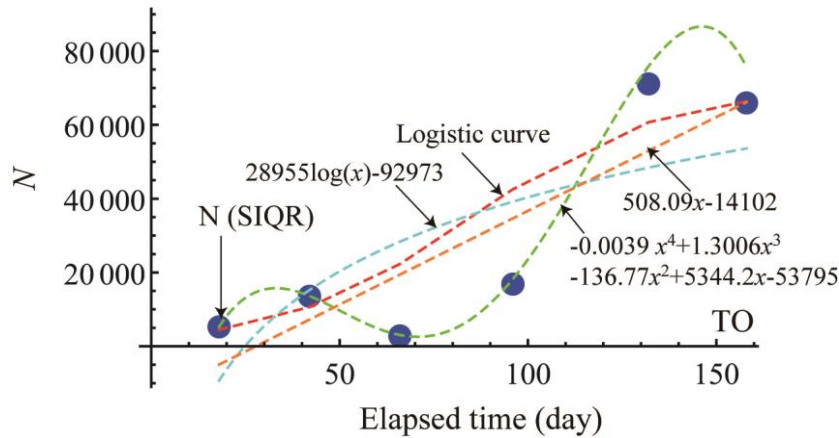

Fig. 9 Identification of the EIOP and simulation of $N(t)$.

As shown in Table 1, the EIOP $N$ has increased significantly from the first wave to the second wave, but it is reasonable to think that it is changing continuously. Fig. 9 shows the EIOP $N$ identified for each section by dividing one wave into three sections using data from Tokyo. $N$ is increasing at the peak of infection. In order to fit the curve, coefficients other than $N$ are calculated as explanatory variables, so $N$ can be obtained from the result of the combination. Therefore, although the change of only $N$ is not calculated, it can be imagined that $N$ generally draws a curve as shown in Fig. 9. Since the calculation of the equation is an iterated integral, this curve can be used as it is for the calculation, but since this curve is also determined by the curve fit, it is convenient to replace it with some function. Fig. 9 shows four types of approximate curves. Considering the sociological aspect of the nature of the infection, if human daily behavior is repeated within a limited range, approximation by growth curve is considered appropriate. Here, we chose a logistic curve for which the analytical solution is known.

## 2.3 SIQR model extension

In order to apply the Effective Infection Opportunity Population (EIOP) described in Sections 2 and 3 to the mathematical model of infection, we extend the SIQR model and consider the effective SIQR model shown in Eq. (1). Eq. (1a) refers to the sum of the term of uninfected person $S$ in the SIQR model equation and the rate of change in the EIOP. The EIOP $N(t)$, which changes with time, can be expressed by Eq. (2) using a logistic curve. Since the increase in the EIOP is added to the term of uninfected person $S$ in Eq. (1a), the decrease in $S$, which acts as a brake on infection, slows down. Furthermore, since the infection coefficient $\beta$ in the equation is $\beta = \beta' / N$, it changes as $N(t)$ changes. Therefore, the $N(t)$ used in the effective SIQR model equation is different from the $N$ identified from the SIQR model equation. Therefore, in order to fit the curve in the effective SIQR model, it is necessary to obtain the coefficient of Eq. (2) again.

Here we use the following symbols:

- $S$ : Uninfected or susceptible,
- $I$ : In-the-city infectant,
- $R$ : Recovered (+ dead),
- $Q$ : Isolated infected (hospitalized + healer),
- $\beta$ : Infection coefficient, $\beta=\beta'/N$   $\beta'$ : Infection rate,
- $N(t)$: Effective Infection Opportunity Population (EIOP),
- $\gamma$: Recovery rate of in-the-city infectant, $1/\gamma$: Healing period (infection period),
- $q$ : Isolation rate,
- $q'$: PCR test isolation rate,
- $\gamma'$: Isolation patient cure rate, $1/\gamma'$ : Healing period.

The effective SIQR equation is given by

$$\frac{dS(t)}{dt} = -\beta S(t)I(t) + \frac{dN(t)}{dt} \tag{1a}$$

$$\frac{dI(t)}{dt} = (1-q')\beta S(t)I(t) - qI(t) - \gamma I(t) \tag{1b}$$

$$\frac{dQ(t)}{dt} = q'\beta S(t)I(t) + qI(t) - \gamma' Q(t) \tag{1c}$$

$$\frac{dR(t)}{dt} = \gamma I(t) + \gamma' Q(t) \tag{1d}$$

$$N(t) = S(t) + I(t) + Q(t) + R(t) \tag{1e}$$

$$N(t) = \frac{b'}{1+c'e^{-a't}} \tag{2}$$

An example of the function in Eq. (2) is illustrated in Fig. 10:

As with the conventional SIQR model, the Euler method was used for the iterated integral, and the Excel 2013 solver was used for the curve fit.

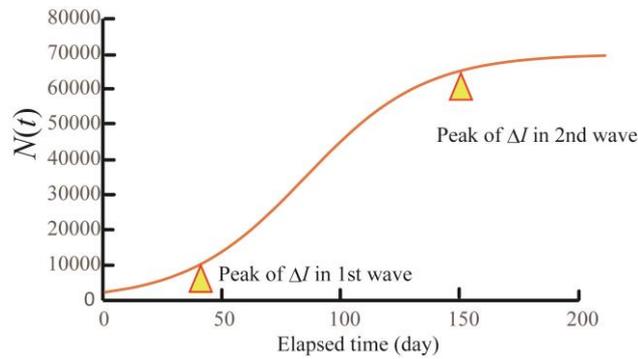

Fig. 10 Image of a function representing $N(t)$.

## 2.4 Prediction of the third wave by the effective SIQR model

Fig. 11 shows the fitting by calculating the effective SIQR model using the data of the first wave and the data up to the middle of the second wave (September 12) in Tokyo. The effective SIQR model is able to continuously regress the first wave and the second wave as expected. The gray curve $N(t)$ shown in the upper left figure is the changing Effective Infection Opportunity Population (EIOP) identified by this calculation. The blue curve $\Delta N$ shown in the upper right is the increment of $N(t)$. The parameters $\beta'$, $q'$, $I(0)$ of the effective SIQR model equation, and the parameters $a'$, $b'$, $c'$ in the curves representing $N(t)$ in equation (2) are obtained as explanatory variables. The blue line drawn in the upper left figure is the EIOP $N$ for each of the first and second waves identified by the conventional SIQR model.

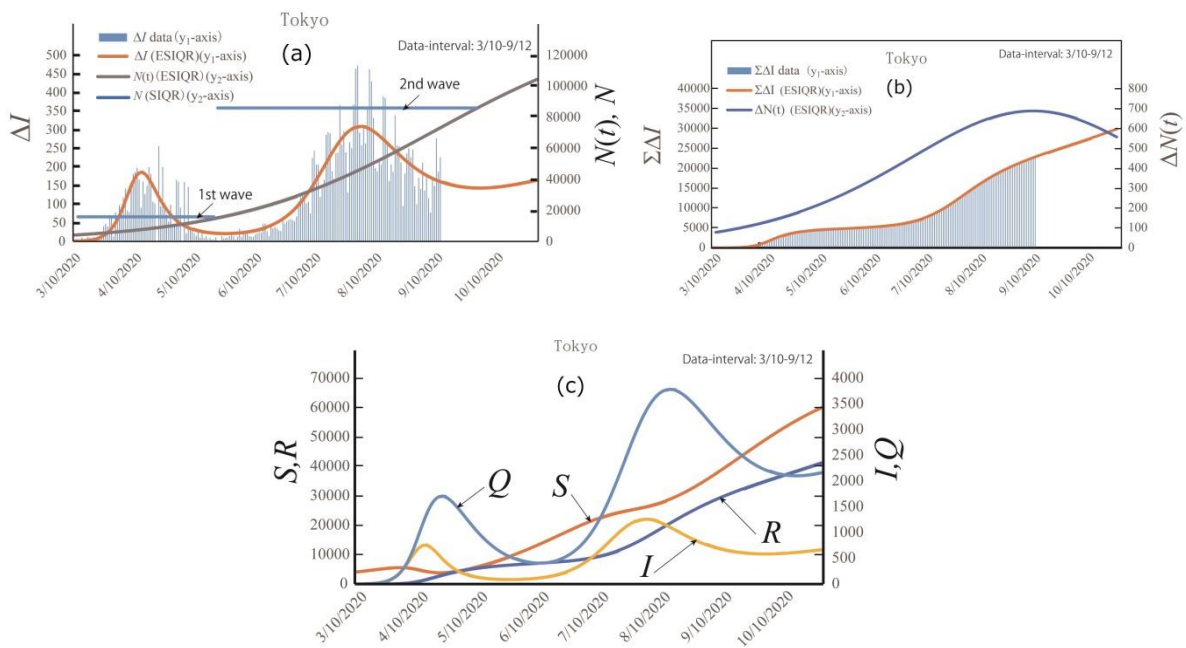

Fig. 11 Fitting with data from 3 / 10-9 / 12 (fitting of 1st and 2nd waves).

Fig. 12 shows the result of fixing the *N* (t) obtained in Fig. 11 and fitting with the data from March 10 to May 21 of the first wave. For comparison, the calculated values in Fig. 11 are shown by brown lines in Fig. 12. It can be seen that the rise of the second wave can be expressed only by the data up to the time of convergence of the first wave alone. However, unfortunately, since there is no data for the second wave when the first wave converges, the curve of *N* (t) shown in Fig. 12 cannot be identified at that time.

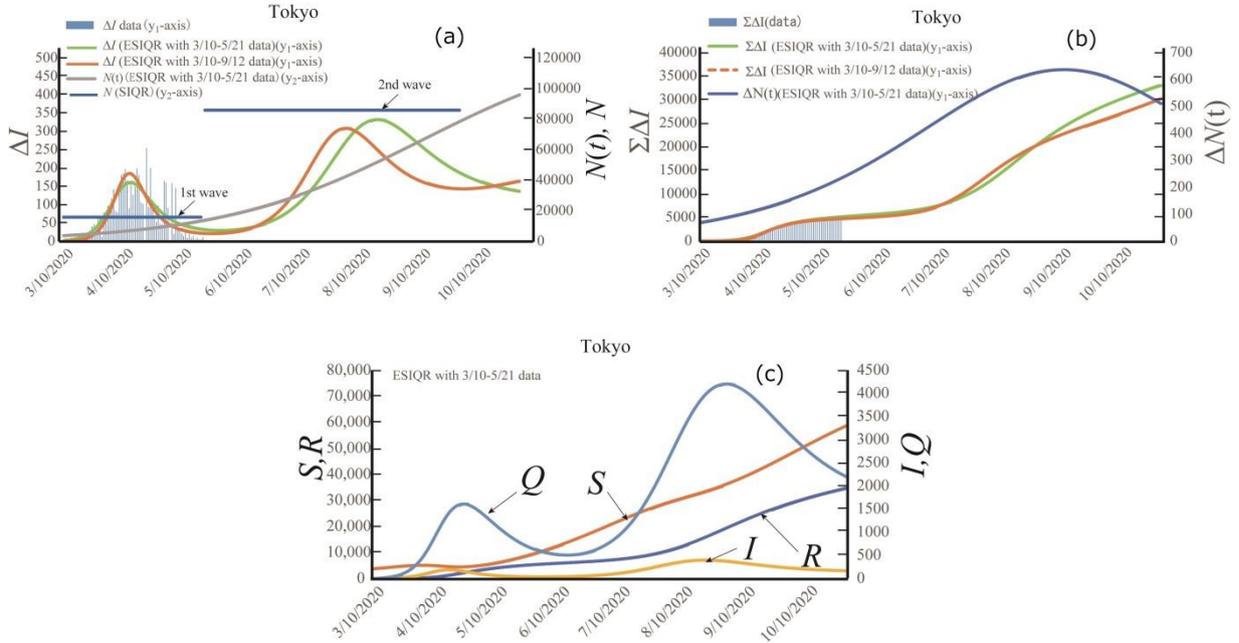

Fig. 12 Fitting with data from 3 / 10-5 / 21 (2nd wave prediction from 1st wave data).

Fig. 13 shows the effective reproduction number *Re'* calculated from the calculated values of the effective SIQR model in Fig. 11. In addition, the effective reproduction number *Re* calculated from the data of the actual newly infected person is described. The number of effective reproductions was calculated by a simple formula according to Toyo Keizai ONLINE [5], which used the data in the calculation of this paper too. Since the effective reproduction number *Re* calculated from the actual data is calculated by increasing or decreasing on a weekly basis, it varies considerably as shown in Fig. 13. Therefore, it may be difficult to judge daily whether the infection is converging or spreading. On the other hand, since the effective reproduction number Re' calculated from the calculated value is stable, it is easy to judge when the infection that crosses the line where *Re'* is 1 is converging or spreading. However, since the prediction curve has been calculated, it is considered that there is no merit in calculating the effective reproduction number *Re'*, but it is confirmed by calculating *Re'* that there is no error in the calculation. Furthermore, the prospect of monitoring is improved when compared with the effective reproduction number *Re* calculated from the data.

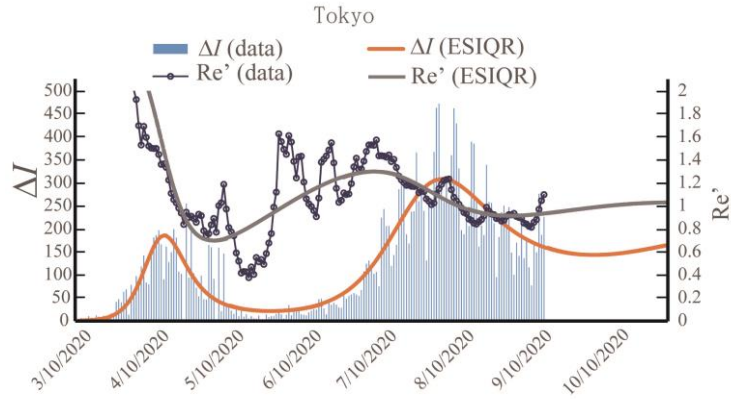

Fig. 13 Calculation of effective reproduction number.

The second wave is still in the process of convergence, but suppose that the third wave occurs on the extension of the $N(t)$ curve identified by the first and second waves. Then, we should be able to predict the third wave using this $N(t)$ curve. Fig. 14 shows the calculation results of an attempt to predict the third wave by fitting with the data of the second wave using this curve of $N(t)$. In this example, although there are signs of the spread of the infection, the third wave has not been launched by the end of November. Whether the EIOP $N(t)$ identified from the data up to the second wave can be effectively used up to the third wave? Alternatively, it is necessary to wait for future data as to whether it is better to use $N(t)$ identified in both the first wave and the second wave. Since we tried the prediction calculation in advance, we think that we are preparing to take countermeasures depending on the result.

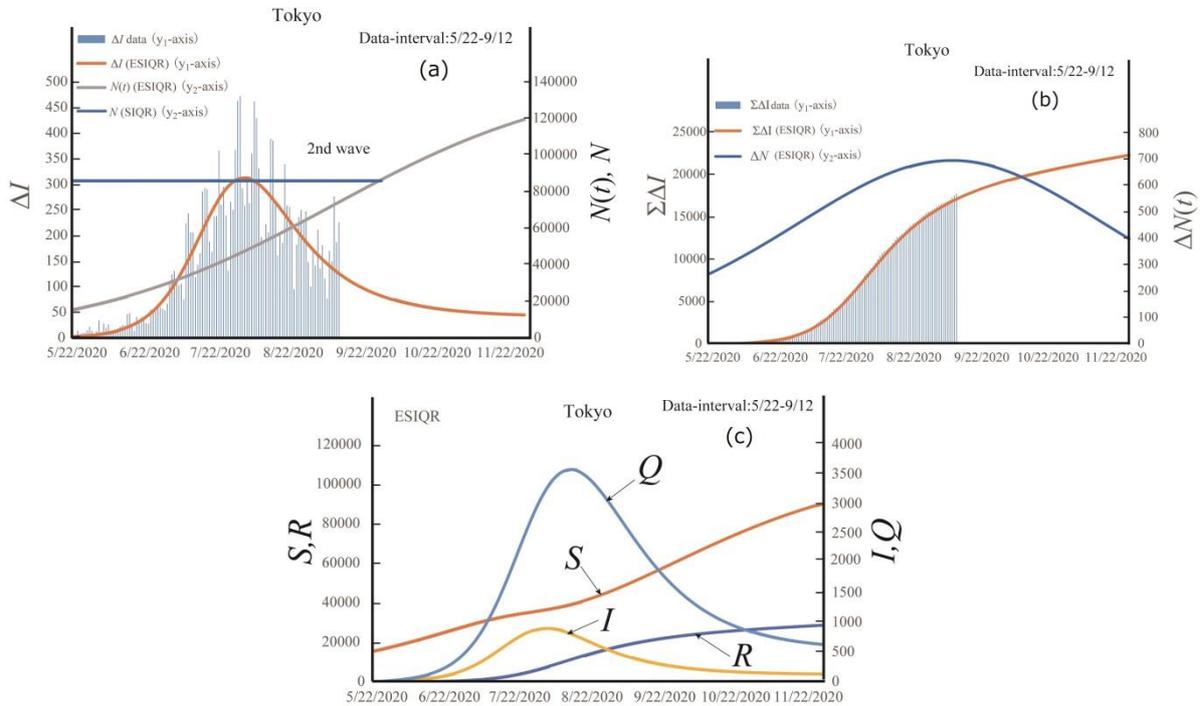

Fig. 14 Fitting with the data of the 2nd wave (5/22-9/12) (3rd wave prediction from the data of the 2nd wave).

It is considered that the results of nationwide self-restraint, masking, and three-closeness avoidance actions taken through the first and second waves are reflected in the $N(t)$ obtained here. Fig. 15 shows a simulation calculation assuming that economic activity becomes more active and the rate of increase in $N(t)$ increases beyond the data of the second wave (after September 12). In this example, the infection begins to increase in mid-October.

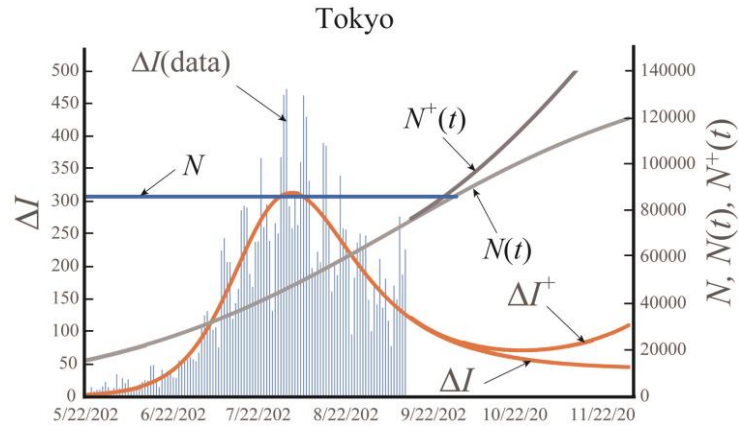

Fig. 15 When the increase rate of the EIOP increases.

Unfortunately, we have not grasped the relationship between economic activity and changes in $N(t)$. So far, we have only shown that the effective SIQR model can be used to simulate changes in infection by changing $N(t)$ in the middle.

The following predictions are possible as a result of calculation using the data up to September 12 with the effective SIQR model. If the third wave occurs, the calculation results in this paper suggest that it will start in mid-October.

Fig. 16 shows the calculation of the second wave forecast using the conventional SIQR model, which was used for the "COVIT-19 Forecast [5]". In the SIQR model, the isolated infected persons (hospitalization + medical treatment persons) Q are calculated. By multiplying this calculated value by the ratio of inpatients and the ratio of severely ill patients in each region, the inpatients and severely ill patients can be predicted. As shown in Fig. 16, the predictive calculation seems to be sufficiently accurate in practice. Fig. 16 is an example of Tokyo. The results of major prefectures and others shown in Table 1 are also calculated and published in the "COVID-19 Forecast", showing good agreement with the actual data. It is useful for medical personnel to be able to predict the number of hospitalized and severely ill patients in this way.

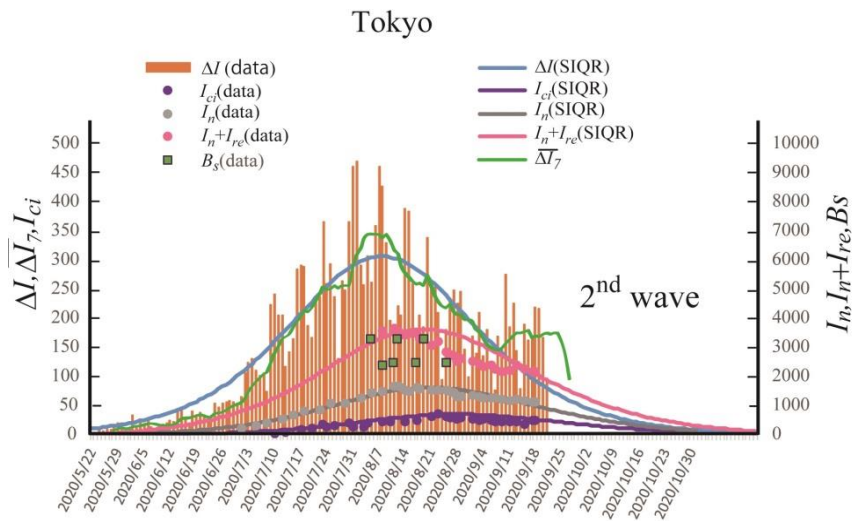

Fig. 16 Calculation of the second wave using the conventional SIQR model.

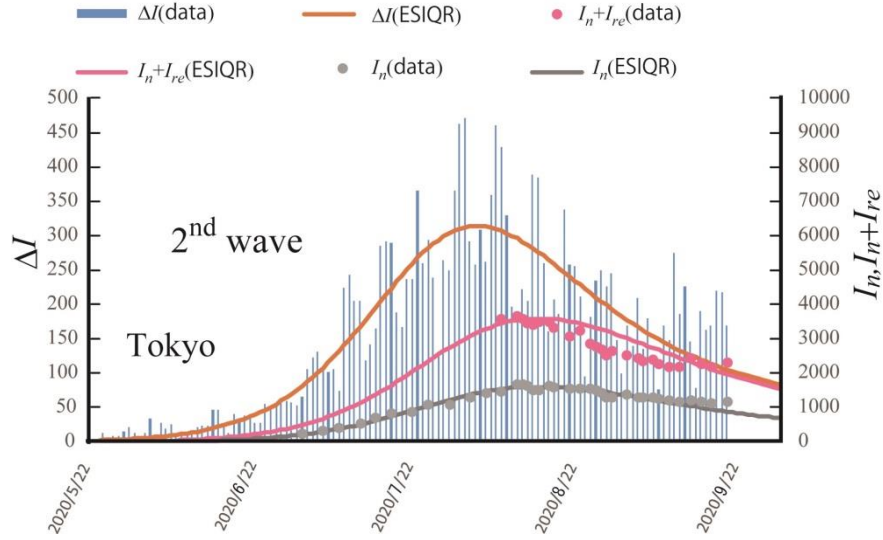

Fig. 17 Calculation of the second wave using the effective SIQR model.

Fig. 17 shows the calculation results and actual data of hospitalized + recuperators and inpatients using the effective SIQR model proposed here for comparison with the calculation using the conventional SIQR model. Predictive calculations equivalent to the conventional SIQR model are possible.

As mentioned above, in the SIQR model, the calculated value of the infected person approaches 0 as the infection converges. Therefore, the SIQR model cannot express the phenomenon that infection continues sporadically after the convergence of one wave. Figs. 16 and 17 are graphs up to September 30, which are almost the same up to this point, but there will be differences in the future.

In order to avoid confusion, the color of the vertical bar of the newly infected person in the figure and the curve of the calculated value is changed between the conventional SIQR model and the effective SIQR model.

## 3. Theoretical introduction of social effects on infection [7]

The uninfected population of the n + 1 wave is smaller than the uninfected population of the nth wave. From that point of view, it should be considered that the infected population of the n + 1 wave is less than the infected population of the nth wave in common sense. On the data observed in Japan, the second wave is about twice as large as the first wave. However, considering the difference in the number of PCR tests, there is an impression that the second wave might be smaller.

Based on the above discussion, it is possible to think of the effective uninfected person S as follows:

$$\frac{dS}{dt} = -\frac{\beta}{N} SI - X, \quad (3)$$

where, X is a social factor. As X, we could ssume

$$X = \alpha \frac{dI_{new}}{dt}. \quad (4)$$

When people see the newly infected population $I_{new}$ increase (that is, $dI_{new}/dt > 0$), they become eager about self-restraint from fear and the effective $S$ decreases. In addition, when $I_{new}$ begins to decrease (that is, $dI_{new}/dt < 0$), effective $S$ and $\beta$ increase due to relaxation. Therefore, the effective reproduction number exceeds 1, and the number of infected persons increases (lack of self-restraint effect).

When the cluster is completely contained, the effect is outstanding, but if even one person misses it, the effect of containment is only a time delay. Therefore, the social factor $X$ is very important. Fig. 18 shows that waves are generated due to changes in social factors.

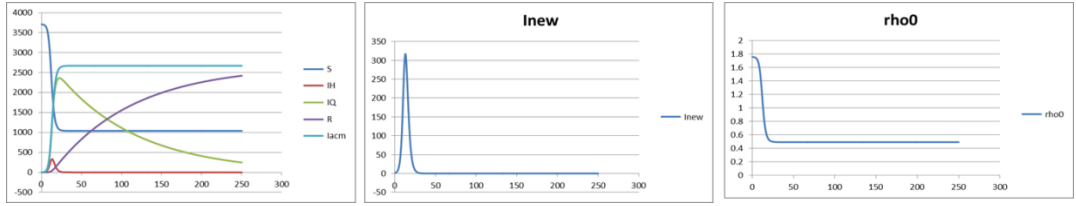

Without social factors (theoretical calculation)

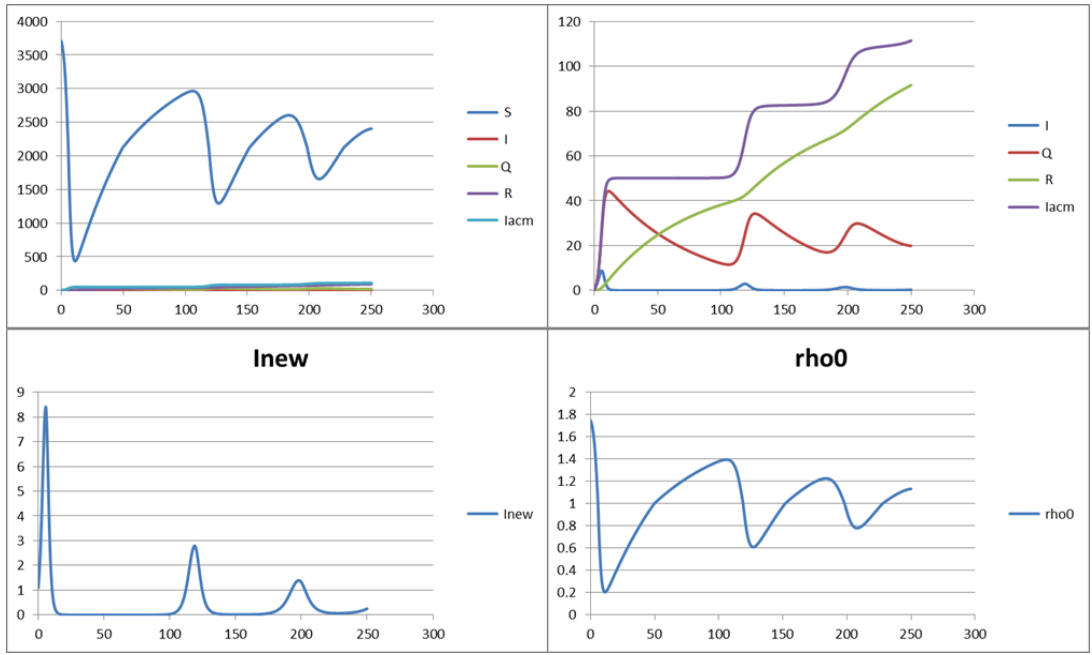

With social factors (theoretical calculation)

Fig. 18: Simulation results by SIQR equation (including self-restraint effect such as mask).

## 4. Extension of SIR-type equation [8]

The system of the SIR infection equation is a nonlinear simultaneous ordinary differential equation:

$$N = S + I + R, \tag{5}$$

$$\frac{dS}{dt} = -\frac{\beta}{N} SI, \tag{6}$$

$$\frac{dI}{dt} = \frac{\beta}{N} SI - \gamma I, \tag{7}$$

$$\frac{dR}{dt} = \gamma I, \tag{8}$$

where $N$, $S$, $I$, and $R$ refer to the numbers of the total population, the uninfected (or susceptible strictly speaking)

people, the infected people, and the recovered people (including dead), respectively, and $\beta/N$ and $\gamma$ the infection coefficient and recover rate, respectively. Although the first terms on the right-hand sides of Eqs. (6) and (7) are not divided by $N$ usually, $\beta/N$ should be used in order to reduce the dependency on the number of the population.

Let the whole space $\Omega$ be divided into the sub-spaces (or cell) $\Omega_i$ ($i = 0, 1, \ldots, I$-1), and $N_i$, $S_i$, $I_i$, and $R_i$ be the number of the population, the number of the uninfected people, the number of the infected people, and the number of the recovered people in sub-space $\Omega_i$, respectively. $\beta_i/N_i$ and $\gamma_i$ are the infection coefficient and recovery rate. The system of the SIR infection equations given by Eqs. (5)-(9) could be extended as follows:

$$N_i = S_i + I_i + R_i, \tag{9}$$

$$\frac{dS_i}{dt} = -\frac{\beta_i}{N_i} S_i(t) \sum_{j=0}^{J-1} \lambda_{ij} I_j \left( t - \frac{r_{ij}}{v_{ij}} \right), \tag{10}$$

$$\frac{dI_i}{dt} = \frac{\beta_i}{N_i} S_i(t) \sum_{j=0}^{J-1} \lambda_{ij} I_j \left( t - \frac{r_{ij}}{v_{ij}} \right) - \gamma_i I_i, \tag{11}$$

$$\frac{dR_i}{dt} = \gamma_i I_i. \tag{12}$$

where $r_{ij}$ are the distance between the position $i$ and $j$, and $v_{ij}$ is the average moving velocity of persons. $\lambda_{ij}$ is a parameter that adjusts the infection of from $j$ to $i$ where $\lambda_{ii} = 1$. Since the infection such as COVID-19 takes place with humans as media of transmission, the infection from $j$ to $i$ is considered proportional to the product of $S_i$ and $I_j$.

In case of a continuous space, the number of the uninfected people, the density of the infected people, and the number of the recovered people are referred to as $s(\mathbf{x})$, $i(\mathbf{x})$, and $r(\mathbf{x})$, respectively, where $\mathbf{x}$ is the coordinates. The system of the SIR infection equations given by Eqs. (5) through (8) could be extended as follows:

$$N = \int_\Omega s(\mathbf{x})d\mathbf{x} + \int_\Omega i(\mathbf{x})d\mathbf{x} + \int_\Omega r(\mathbf{x})d\mathbf{x}, \tag{13}$$

$$\frac{\partial s(\mathbf{x},t)}{\partial t} = -\frac{\beta'(\mathbf{x},t)}{\rho(\mathbf{x})} s(\mathbf{x},t) \int_\Omega \lambda(\mathbf{x},\xi) i\left(\xi, t - \frac{r(\mathbf{x},\xi)}{v(\mathbf{x},\xi)}\right) d\xi, \tag{14}$$

$$\frac{\partial i(\mathbf{x},t)}{\partial t} = \frac{\beta'(\mathbf{x},t)}{\rho(\mathbf{x})} s(\mathbf{x},t) \int_\Omega \lambda(\mathbf{x},\xi) i\left(\xi, t - \frac{r(\mathbf{x},\xi)}{v(\mathbf{x},\xi)}\right) d\xi - \gamma(\mathbf{x},t)i(\mathbf{x},t), \tag{15}$$

$$\frac{\partial r(\mathbf{x},t)}{\partial t} = \gamma(\mathbf{x},t)i(\mathbf{x},t), \tag{16}$$

where $\rho(\mathbf{x})$ is the density of the population at the coordinates $\mathbf{x}$, and $\beta'(\mathbf{x},t)$ has the dimension of [Area$^{-1}$Time$^{-1}$].

For simplicity, if we consider that the infection occurs between neighboring cells alone, a following system of the simplest infection equations in one-dimensional space is considered:

$$N_i = S_i + I_i + R_i, \tag{17}$$

$$\frac{dS_i}{dt} = -\frac{\beta_i}{N_i} S_i(t) \frac{1}{w_0 + w_1 + w_2} \left( w_0 I_{i-1}(t) + w_1 I_i(t) + + w_2 I_{i+1}(t) \right), \tag{18}$$

$$\frac{dI_i}{dt} = \frac{\beta_i}{N_i} S_i(t) \frac{1}{w_0 + w_1 + w_2} \left( w_0 I_{i-1}(t) + w_1 I_i(t) + + w_2 I_{i+1}(t) \right) - \gamma_i I_i, \tag{19}$$

$$\frac{dR_i}{dt} = \gamma_i I_i.\tag{20}$$

The boundary condition is given by

$$I_{-1} = I_1, \quad I_I = I_{I-2}.\tag{21}$$

Firstly, the appropriateness of the above-extended SIR equation is checked. If the uninfected and infected people are distributed uniformly, the case can be solved both by the SIR and extended SIR equation. It is ascertained that the same result is obtained by both methods as shown in Fig. 19. The input data for this case is shown below:

For SIR theory

$$N_T=400, \ S(0)=395, \ I(0)=5, \ R(0)=0, \ \beta=0.4, \ \gamma=0.08, \ dt=0.1.\tag{22}$$

For extended SIR theory applied to the equal partition of the space

$$N_T=400, \ I=25, \ N_i=16, \ S_i(0)=(N_T-5)/I=395/25, \ I_i(0)=5/I, \ R_i(0)=0,$$
$$\beta=0.4, \ \gamma=0.08, \ [w_0,w_1,w_2]= [1,0,1] \text{ or } [1,1,1] \text{ or } [1,10,1], \ dt=0.1.\tag{23}$$

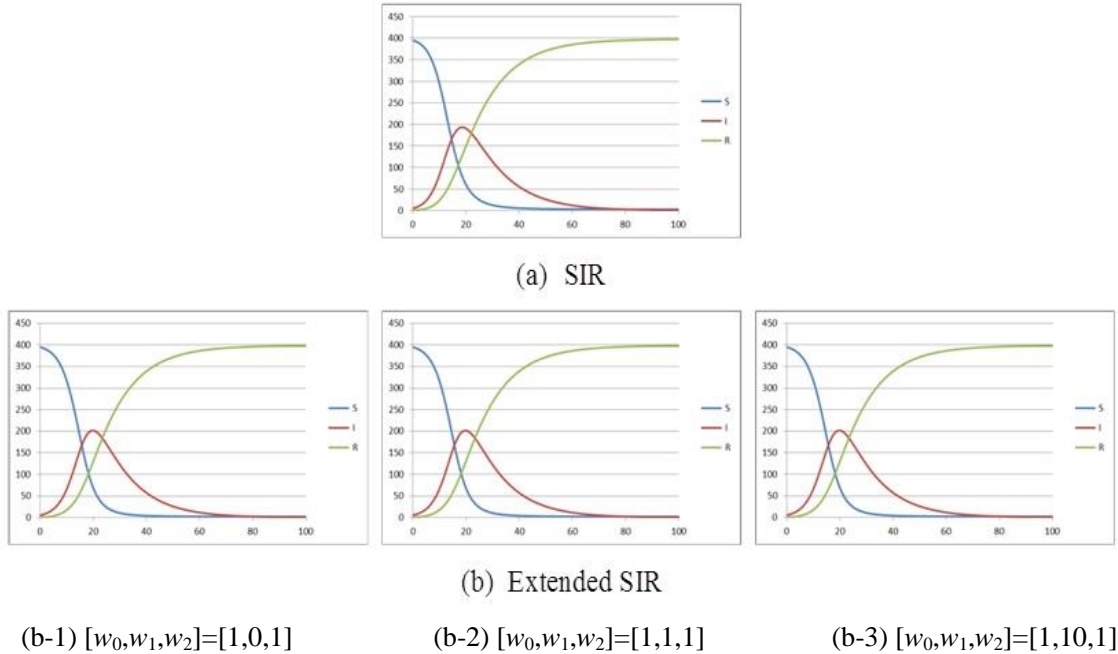

(a) SIR

(b) Extended SIR

(b-1) $[w_0,w_1,w_2]=[1,0,1]$      (b-2) $[w_0,w_1,w_2]=[1,1,1]$      (b-3) $[w_0,w_1,w_2]=[1,10,1]$

Fig. 19. Comparison of numerical results between SIR method and extended SIR method

(Horizontal axis: Time, Vertical axis: Number of people).

Next, the calculations where the initial uninfected people are equally distributed, but the initial infected people are concentrated at the center of the population are conducted by SIR and extended SIR methods, and the results are compared. The input data are as follows:

For the SIR method

$$N_T=400, \ S(0)=395, \ I(0)=5, \ R(0)=0, \ \beta=0.4, \ \gamma=0.08, \ dt=0.1.\tag{24}$$

For the extended SIR method (equally partitioned)

$$N_T=400, \ I=25, \ N_i=16, \ S_{11}(0)=15, \ S_{12}(0)=13, \ S_{13}(0)=15, \ I_{11}(0)=1, \ I_{12}(0)=3, \ I_{13}(0)=1,$$
$$\text{unspecified } S_i(0), \ I_i(0), \text{ and } R_i(0) \text{ are all zero},$$
$$\beta=0.4, \ \gamma=0.08, \ [w_0,w_1,w_2]=[1,10,1], \ dt=0.1.\tag{25}$$

As shown in Fig. 20, the spreading of infection is represented interestingly.

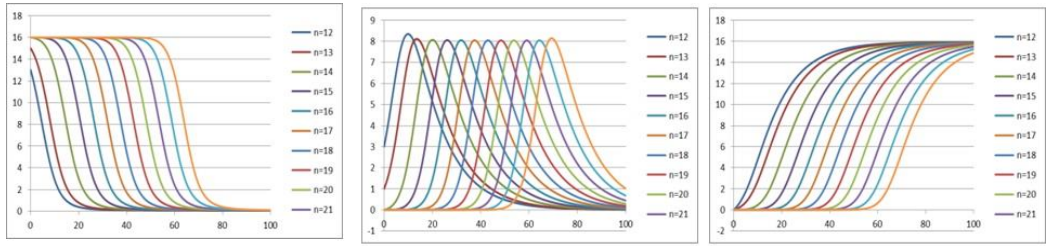

Uninfected people *S*  Infected people *I*  Recovered people *R*

(a) Temporal distribution

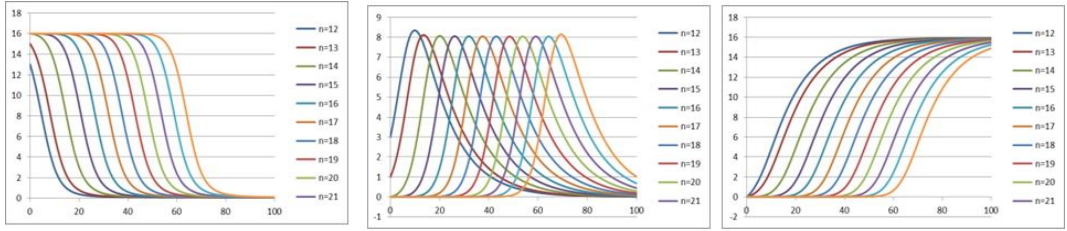

Uninfected people *S*  Infected people *I*  Recovered people *R*

(b) Spatial distribution

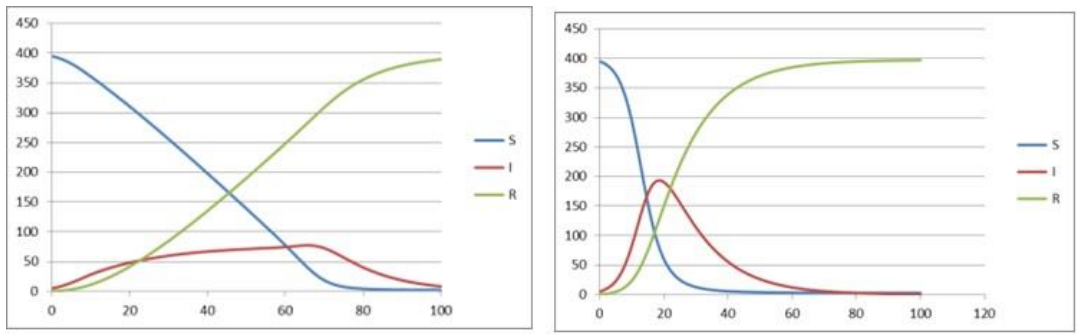

(c) Extended SIR (total)  (d) Original SIR

Fig. 20. Numerical results by Extended SIR method: Spreading of infection

(Horizontal axis: Time, Vertical axis: Number of people).

When the weight $w_0$, $w_1$, $w_2$ is changed, the spreading of the infection changes as shown in Fig. 8.

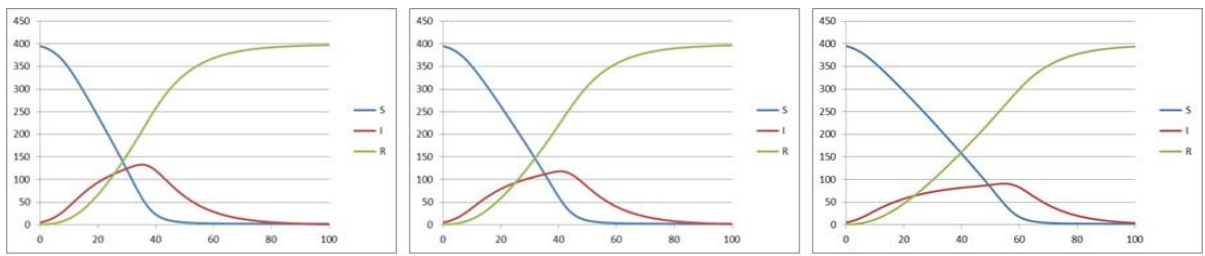

(a) [$w_0$,$w_1$,$w_2$]=[1,0,1]  (b) [$w_0$,$w_1$,$w_2$]=[1,1,1]  (c) [$w_0$,$w_1$,$w_2$]=[1,5,1]

Fig. 21. Numerical results by Extended SIR method: Effects of weight $w_0$, $w_1$, $w_2$

(Total, Horizontal axis: Time, Vertical axis: Number of people).

## 5. Conclusion

Prediction is the first step in judging, making decisions, and taking action. Even in the case of COVID-19 infection, if you skip the prediction and take the first step, you will fail. Unfortunately, for COVID-19 infections, even with the release of daily updated data, we have not yet been exposed to reports that explain what will happen next week on a scientific basis.

It is clear that if the third wave can be predicted accurately, it will be useful for the judgment of the government. Here, using the proposed effective SIQR model, we showed the possibility of continuously simulating two or more waves by changing the Effective Infection Opportunity Population (EIOP). As a result, there are still issues such as at what stage the EIOP $N(t)$, which is the basis of the effective SIQR model calculation, can be estimated accurately, but we believe that we have reached the entrance to the third wave prediction.

As mentioned in this paper, the EIOP and other parameters used in the calculations are influenced not only by biological properties but also by social science behavior, so it is difficult to estimate from physical grounds alone. Currently, the SIR model that can be used to identify data on new coronavirus infections is the mean-field model. The phenomenon that actually occurs is a non-mean field that extends from the cluster. Therefore, we have discussed in this paper that even if the mean-field model is fitted using conventional data fitting, satisfactory results cannot be obtained. Therefore, we introduced the concept of EIOP and used the mean-field model together with it. This is a problem regarding a new data fitting method that makes it possible to simulate the phenomenon of the non-mean field.

Since the method discussed here does not identify the coefficients used in the equation individually, I think that many people have doubts about the physical basis. On the other hand, it is an undeniable fact that the prediction calculations for inpatients and critically ill patients performed by the method discussed in this paper are in good agreement with the actual data. The origin of the prediction begins with finding a curve that fits well with the changing phenomenon. In that sense, the effective SIQR model and its calculation method discussed in this paper are interpreted as providing a curve ruler that applies to two or more consecutive waves of the new coronavirus infection phenomenon.

Data on the movement of people are frequently published, but since infection starts if even one infected person invades a population of uninfected people, the magnitude of the spread of infection is not necessarily determined only in proportion to the movement of people. It is important to consider how movement affects. In the effective SIQR model, the EIOP is sequentially calculated and poured into the formula of uninfected person $S$. Therefore, if the population flowing into the area is known from the program that calculates the movement of people in chronological order, the effective change in the uninfected person $S$ having the risk of infection can be grasped. Thinking in this way, it seems that calculations linked to the movement of people are possible. For example, we believe that the goal of the new corona prediction is to link the human migration program [5] with the mathematical model of infection as described here. This is also an issue since the study of the first wave.

The effective SIQR model can also simulate what Odagaki had already pointed out the effect of isolation of infected people in the city by PCR using the SIQR model. In addition, the effect of the PCR test performed on the first wave on the second wave can be simulated, but it is omitted here.

With the introduction of EIOP, the phenomenon of the non-mean field can be treated by the equation of mean-field to some extent, but the existence of data is indispensable for that purpose, and it cannot be used for theoretical simulation. So the authors came up with two methods to improve the situation. One is a method of theoretically incorporating the social effect on the infectious phenomenon, and the other is spatially discretizing the SIR type equation for dealing with the mean-field so that the phenomenon of the non-mean field can be dealt with directly. We also discussed both of these and obtained useful directions regarding the mathematical analysis of future infection phenomena.

**Acknowledgments**

The authors are deeply grateful to the members of the Mathematical Model Study Group for New-type Coronavirus Infectious Diseases, especially Professor Emeritus Takeshi Kinoshita of the University of Tokyo who accepted the representative, and Takashi Odagaki of Kyushu University who proposed the SIQR model or the basis of our model. We also acknowledge Dr. Ryosuke Yano's big effort in preparing figures and a table.